# INTEGRATION OF WIRELESS SENSOR NETWORKS WITH VIRTUAL INSTRUMENTATION IN A RESIDENTIAL ENVIRONMENT

Grigore STAMATESCU[1], Valentin SGÂRCIU[2]

*This paper presents an approach to integrate wireless sensor networks (WSN) with the LabVIEW graphical development environment through a dedicated software driver. As personal constribution, a system architecture and concept implementation are described, in the context of a smart house monitoring scenario. Data acquisition is performed via the deployed wireless sensor network with focus on three main parameters: temperature, humidity and light. The data logging, monitoring and control functions are realized through a virtual instrumentation project. This also enables an easy-to-use user interface and the accesibility of data through standards-based web server technologies. The potential of remote monitoring and control through mobile terminals is opened up.*

## 1. Introduction

Recent advances in computer science, telecommunications and electrical engineering have converged into the field of wireless sensor networks (WSN) [1]. These tiny embedded networked devices, which consist of processing units, low-power radios, various sensors and a power supply, usually in the form of batteries, can be deployed in an interest area and relay important data back to a base station. There, data is stored, processed and analyzed and can be made available to the user. In a broader view, wireless sensor networks can be seen as essential components of the *Internet of Things* concept [2].

LabVIEW – the graphical programming environment, offers an intuitive way for engineers and scientists to quickly deploy applications for testing, data acquisition and control in the form of virtual instruments (VIs). A VI consists of two parts. The block diagram implements the program logic by wiring together standardized library functions and control structures or user-defined routines. The front panel is the user interface which consists of standard elements such as:

---

[1] Assist., Faculty of Automatic Control and Computers, University POLITEHNICA of Bucharest, Romania, e-mail: grigore.stamatescu@upb.ro
[2] Prof., Faculty of Automatic Control and Computers, University POLITEHNICA of Bucharest, Romania, e-mail: vsgarciu@aii.pub.ro

buttons, graphs, indicators assembled to provide an intuitive view of the program scope.

Two of the most important potential applications of wireless sensor networks are environmental monitoring and home and building automation systems. In the first case, sensor nodes are spread over a certain area (e.g. urban landscape, fields, forest etc.) for monitoring a range of parameters. Various research and commercial WSN platforms have been developed which offer integrated sensors for temperature, atmospheric pressure, humidity and ambient light intensity as well as more specialized systems for air pollutants detection [3], ambient sound or others. Environmental data collection over large areas poses specific issues which have to be dealt with, such as random node placement, battery life and network topology.

On the other hand, in home and building automation systems, nodes can be placed in predefined positions in order to optimize the system efficiency and can be battery but also mains powered. The network topology is usually static. In existing buildings the main benefit of using wireless sensor networks for monitoring the environment is that of low installation and maintenance costs with good performance and scalability. Also, low-power wireless mesh networking has the potential to mitigate coverage issues which usually appear in such situations by relaying data to the base station through multi-hop communication.

In this paper we present the integration of a Memsic wireless sensor network composed of IRIS nodes and MIB520 USB interface with LabVIEW and related technologies with the goal of narrowing the gap between the electrical engineering and computer science research and specialized domain users in designing and operating real WSN applications. This builds upon the work in [2] and [3] but with a more detailed view of the intricacies of the software driver for serial communication with the base station, a more extensive evaluation of the system deployment and the enabling technology behind the distance monitoring component.

Literature offers various approaches to integrating specific wireless sensor networks platforms with the LabVIEW environment. In [4] a technical solution for sensor network monitoring based on virtual instrumentation is presented. The authors describe the sensors as distributed measuring points for physical variables in distributed parameter systems. A generic data aquisition application is described which employs the dedicated LabVIEW driver for Crossbow WSN [5].

The authors of [6] present a more specific application which focuses on the in-depth study of the use of wireless sensor networks to monitor and control a swiftlet habitat. The application handles communication with the sensor network, the program logic and the graphical user interface but implements other functions such as remote access and a database system for record and management purposes. We build on top of these two approaches and devise a system which offers both a variety of functions and scalability and development possibilities.

The paper is structured as follows: Chapter 2 describes the general architecture of the system, highlighting the benefits of using wireless sensor networks in building environment and home automation tasks. Chapter 3 presents the hardware structure configuration considerations and details of the software driver used. Chapter 4 is dedicated to the development and implementation of the virtual instrumentation project along with the challenges encountered and experimental evaluation along with the shared variable engine that enables the distance monitoring. Chapter 5 ends the paper with conclusions and future work.

**2. System Overview**

The starting point in our scenario is that, by leveraging unique wireless sensor networks characteristics, like low-power wireless communication and battery operation, we can obtain high temporal and spatial resolution monitoring of the indoor environment. This would enable us to have fine-grained control over actions such as heating, cooling and lighting with a considerable impact on the energy consumption and utility bills. Our study is applied to a common 1-2 dormitory residential flat. We assign a sensor node equipped with temperature, humidity and light sensors to each room. After the nodes are powered on, they self-organize in a mesh network, start sampling the sensors and transmit data packets over the wireless link. Due to the difficult nature of the indoor environment, phenomena such as fading, multipath reflection and interference caused by walls, large metal objects or high powered devices operating in the 2.4 GHz ISM (Industrial, Scientific and Medical) band may appear. These reduce the radio coverage of the nodes and so, the nodes which fall outside the range of the base station have to rely on neighbors to forward their packages to the root of the newtork. Subsequently, the base station starts receiving two types radio packets: data packets contain the raw sensor values and health packets contain information about the network status. Data is then forwarded via serial communication to the gateway which, in a conventional wireless sensor network, has the essential task of storing and forwarding it to interested third parties and, in specific cases, to process, analyze and display it. In our case, the gateway is a personal computer with a USB connection, running a custom designed virtual instrumentation project which uses a dedicated software driver to parse the serial messages and extract useful data from the byte stream. For our scenario we consider a data rate of 10-60 seconds of sampling and transmission to be reasonable whilst expecting a node lifetime before battery replacement of several months.

The overview of the proposed system architecture is shown in Fig. 1. It is to be noticed that we have also included the possibiliy of actuation in our scenario i.e. the gateway is equipped with a data acquisition board hard-wired to actuators for the heating, A/C and lighting systems of each of the rooms. In this work the feedback control based on simple bipositional control algorithms has been implemented only in simulation and we describe the control logic in Chapter 4. The approach is scalable to larger area indoor monitoring applications at the floor or building level.

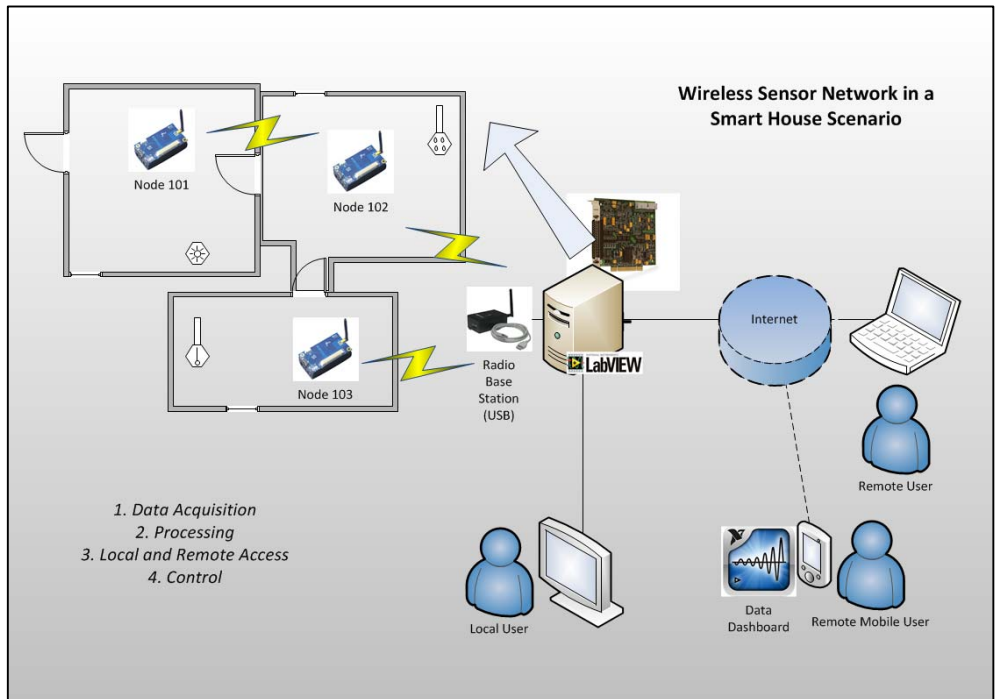

Fig. 1. System Overview

### 3. Hardware and Software Implementation Considerations

Hardware used is commercially available from Memsic Inc. and consists of IRIS sensor nodes and an USB radio base station. The IRIS node has a modular structure, comprised of a radio/processing board, a sensor board, compact plastic enclosure and 2 AA batteries. The XM2110 [7] is the main radio/processing board, which hosts an ATMega 1281 8 bit MCU and a IEEE 802.15.4 compliant RF230 radio transceiver operating in the 2.4GHz ISM band. This is the newest module in the line of the original Berkeley motes and is supported by the open-source community under TinyOS 1.x and 2.1 – an event-based, low footprint operating system for resource constrained devices. Compared to the previous iteration – MicaZ, the producer mentions better performance in terms of radio coverage and improved energy efficiency. The MTS400 sensor board [8] connects to the main board through the 51-pin connector and offers the following sensors:
- humidity and temperature sensor (Sensirion SHT11);
- barometric pressure and temperature sensor (Intersema MS55ER);
- light sensor (TAOS TLS2550);
- 2-axis accelerometer (Analog Devices ADXL202JE).

The radio base station is made up of a IRIS radio/processor board connected to a MIB520 USB interface board via the 51-pin expansion connector. The interface board uses a FTDI chip and provides two virtual COM ports to the host system. COM$x$ is used for programming the connected mote and COM$x+1$ is used by middleware applications to read serial data.

Fig. 2 displays the hardware used to carry out the experimental part of this paper, in a laboratory setting. Throughout the experiments, three nodes are programmed with IDs 101, 102 and 103 while the base station has ID 0. Radio communication is set to channel 26 (2480 MHz center frequency) of the IEEE 802.15.4 defined spectrum allocation, in order to minimize interference by high-powered radio devices such as WiFi access points. Motes can be configured in either HP – High Power mode or in LP – Low Power. In HP, the node MCU and radio are always powered on while LP uses aggressive duty-cycling strategies, that keep radio activity to a minimum and maximize battery life with the drawback of longer network set-up times and higher end-to-end latency. For the experiments described in this paper, nodes are programmed with a HP firmware version that enables faster debugging and accelerated data analysis for our virtual instrumentation project. While this would be unfeasible in a real-world deployment due to the expected days to weeks lifetime of the batteries compared to the months to years lifetime in the low-power version, the framework for WSN integration with virtual instrumentation described below works with both approaches. Table 1 presents, as a reference, the current draw for the IRIS mote under various modes of operation:

*Table 1*

**Current Requirements for the IRIS Board in Different Operating Modes**

| [mA] | Pr. full | Pr. slp | Rad. rcv | Rad. tr | Rad. slp | Flsh wr | Flsh rd | Flsh slp |
|---|---|---|---|---|---|---|---|---|
| **IRIS** | 8 | 0.008 | 16 | 17 | 0.001 | 15 | 4 | 0.002 |

Nodes run a custom low-power networking layer called XMesh, implemented as module firmware in TinyOS 1.x. This enables the motes to self-organize in a wireless mesh network and relay sensor values in a reliable fashion to the base station.

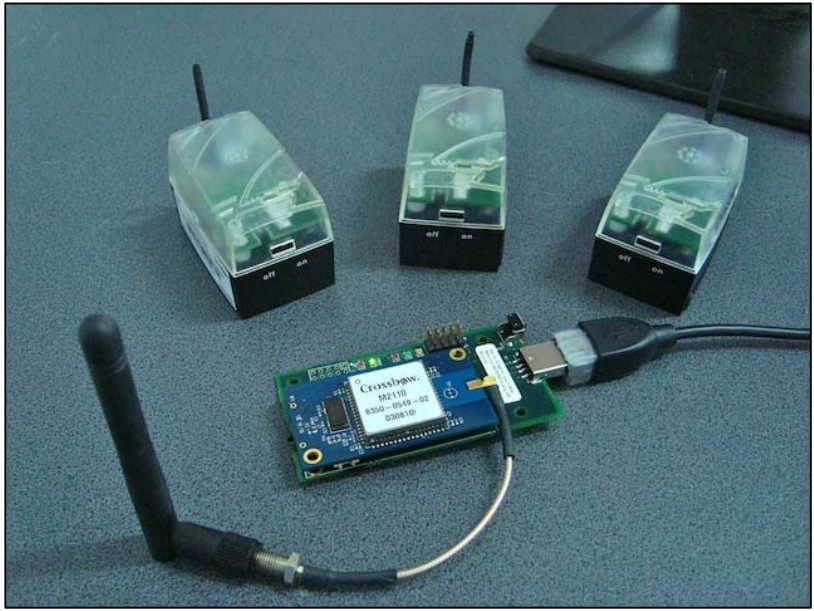

Fig. 2. Memsic IRIS WSN Nodes and USB Radio Base Station

In our particular case, for the MTS400 sensor board, the radio data packet has the structure depicted in Fig. 3.

| Bytes: 5 | 0/7 | 4 | 2 | 2 | 2 | 20 | 2 |
|---|---|---|---|---|---|---|---|
| TinyOS Header | XMesh Header | XSensor Header | Voltage | Humidity | Temperature | | CRC |
| | | | MTS400 Payload | | | | |

Fig. 3. XMesh Message Structure for the MTS400 Sensor Board [11]

The message starts with a compulsory TinyOS header of 5 bytes. A 0-7 bytes XMesh header follows with routing layer data such as: source address, origina address, sequence number and application ID. The XSensor header includes information about the sensor board ID, packet ID and parent node. The last 2 bytes of the message contain the the Cyclic Redundancy Check (CRC) error-detecting code. Within the scope of our application, the "MTS400 Payload" consists of 26 bytes containing the data from the sensors. This data has to be extracted from the message structure and raw numeric values have to be converted to engineering units. As an example, computing the temperature in degrees Celsius from the reading of the SHT11 humidity and temperature sensor [12], involves the folowing computation:

$$T = d_1 + d_2 \cdot SO_T \qquad (1)$$

where T is the temperature in Celsius or Fahrenheit degrees and $SO_T$ is the raw numeric value. The temperature conversion coefficients $d_1$ and $d_2$ to obtain a temperature in Celsius degrees, for a supply voltage of 2.5V and a 14bit resolution are -39.6 and 0.01 respectively. Therefore, the final equation becomes:

$$T = -39.4 + 0.01 \cdot SO_T \qquad (2)$$

The LabVIEW software driver for the chosen suite of hardware modules consists of a LLB library which includes a collection of VIs and subVIs. Each one implements specific routines to enable the serial connection to the interface board, start listening for data on the serial bus, reading the serial buffer, interpreting/parsing the data and converting the values to an useful format.

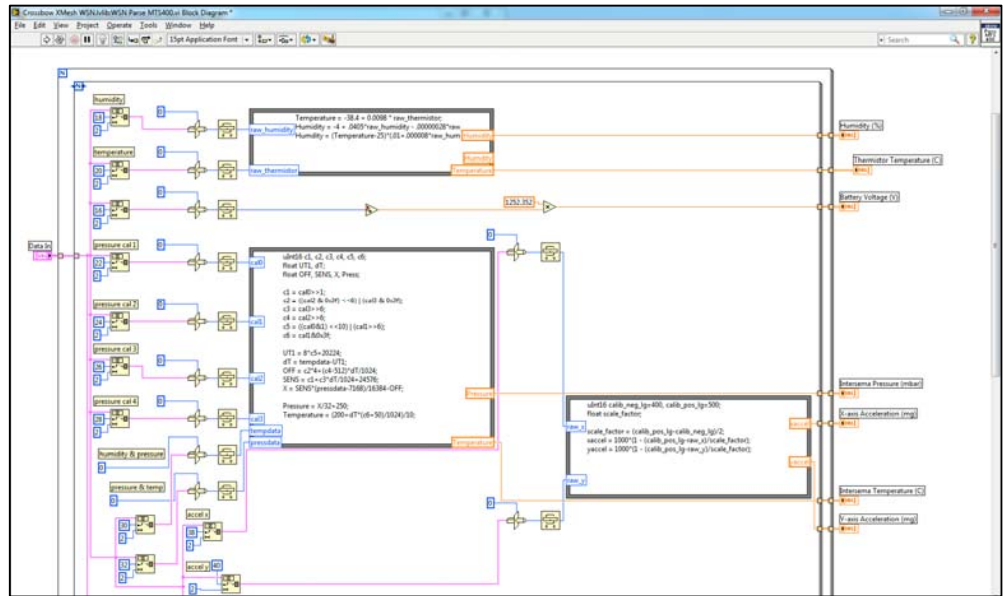

Fig. 4. Block Diagram Fragment of MTS400 Message Parser VI

The message parser subVI in particular processes the input string from serial buffer read function output. The result of the message parsing corresponding to data from the array of sensors installed on the sensor board. Fig. 4 contains a screenshot of the block diagram for the message parsing subVI in the case of the MTS400 sensor board. The logic follows an iterative process inside a series of *for* loops. The data stream is split into useful chunks of bytes according to the message structure and the values are converted to engineering units using their characteristic equations inside MathScript nodes.

Having defined the system architecture and introducing hardware and software components as building blocks of our Smart House monitoring scenario, we go further to describe and evaluate the developed application.

### 4. SmartVI – a Virtual Instrumentation Application to Showcase Wireless Sensor Networks Integration

The goals we have proposed ourselves for the monitoring application reflect in the following tasks that it performs:
- aquires data from the wireless sensor network inside the LabVIEW environment by using specific functions of the Crossbow XMesh driver;
- presents data in an intuitive fashion to a non-specialized end-user;
- displays historical graphs of the monitored parameters (temperature, humitidy, light);
- logs data to specific LVM measurement files for further analysis and processing;
- implements control logic simulation for heating, cooling and lighting devices;
- publishes data, making it available for remote monitoring using shared variables.

The main view of the application is shown in Fig. 5. The user interface is divided into three parts, each one corresponding to a room of our Smart House. Intuitive display elements are placed into each of these frames and display the temperature (degrees Celsius), humidity (%) and light (lux). To observe the historical evolution of the three parameters, we have placed graphs into dedicated tabs which offer the user a global view. The last tab functions as a control panel where the user has the possibility of setting the desired room temperatures and the light activation thresholds.

For running the application, the user has to either have a copy of the LabVIEW environment, it can be built as an executable program and run on client machines through a downloaded runtime or it can be published as a remote panel and be operated through the Internet via a web browser. The cross-platform characteristics of the graphical development system enable it to be run on different host operating systems or in virtualized environments.

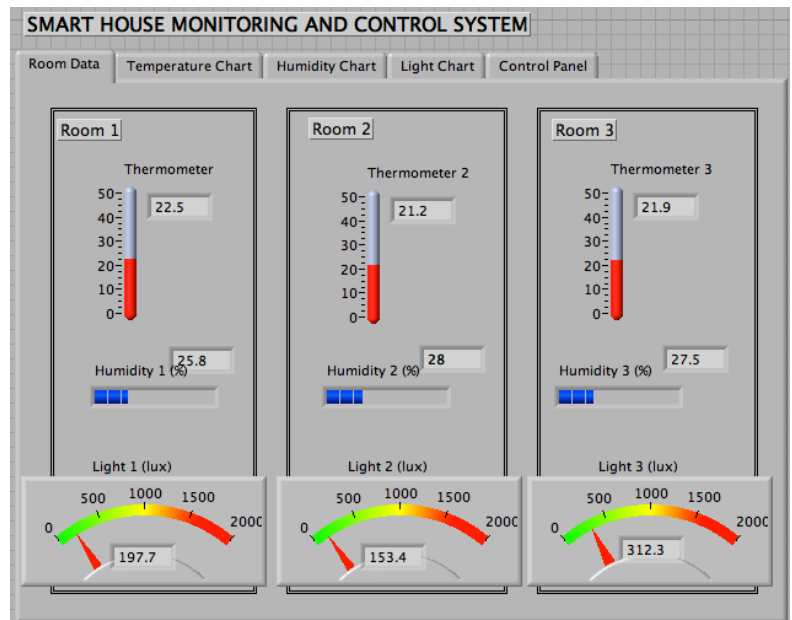

Fig. 5. Front Panel of the SmartVI – Main View

The control logic is as follows: heating/cooling and lighting are treated separately. The system parameters can be tuned in the dedicated tab of the virtual instrument front panel. The heating and cooling bipositional control logic relies on a temperature set-point. If temperature in one of the rooms drops by more than 1 degree, the heating ON signal is activate. When the temperature increases by more than 1 degree, the cooling ON signal is activated. The light control logic checks if the light in one of the rooms drops below a certain threshold (e.g. 200 lux) and if movement has been detected in that room, the light ON signal is activated for that room. The light threshold is user-defined. The control logic panel is depicted in Fig. 6.

To enable the real-world application of this control logic, a suitable data acquisition board has to be installed in the server computer, an the control signals wired to the appropiate actuators. This can be done either through National Instruments hardware which offer good support and deep integration with LabVIEW or through third-party suppliers which offer their own drivers. Conventional, wired actuation is considered the safe way to approach such a control application although researchers have experimented and there are working deployments of wireless low-power actuation systems.

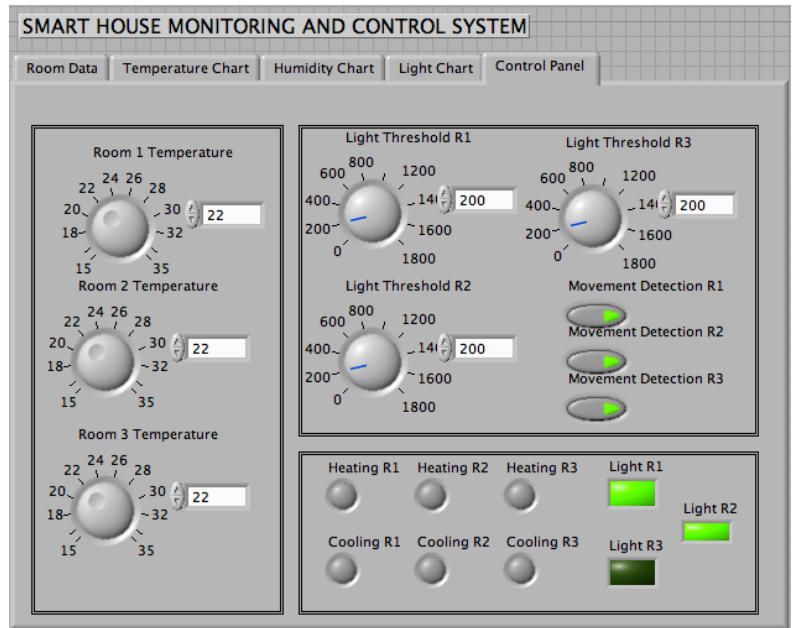

Fig. 6. Front Panel of the SmartVI – Control Panel Tab

The system publishes the room parameters as shared variables to be accesed locally or remotely, through the internet. The shared variable engine abstracts network communication from the LabVIEW block diagram and makes it easy to share data between networked systems [3]. The type of shared variables we are using for our particular scenario consist of networked-published shared variables of of double type. After being added to the project library of a certain LabVIEW application, the shared variables take the form of a terminal which can be dropped onto the block diagram to be written or read from. They also offer error input and output terminals to notify the user of communication errors and a timestamp output that informs about when the value read was written. A specific UDP protocol is impelemented under this abstraction layer to send data to a server called the shared variable engine, which then publishes the data to all clients on the network reading the shared variable.

A block diagram fragment, highlighting the data aquisiton flow from the USB radio base station and the publishing of light information to local and shared variable appears in Fig. 7. Graphical programming differs from standard text-based programming languages in that a block executes whenever data is available to all its compulsory input terminals. This concept is called *dataflow*. In the main loop data from the serial port is parsed by the dedicated function described in the previous section and data is wired to the corresponding local indicators and shared variables. Upon terminating the application through the *Stop* button on the main panel, the data acquisition is halted and the serial connection is closed to be made available for other applications or system services.

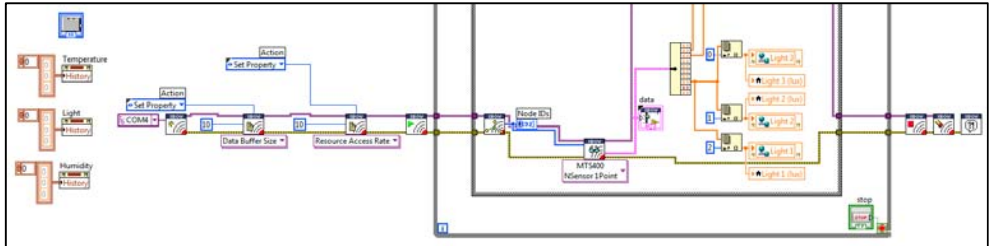

Fig. 7. Block Diagram Fragment of the Main Application Virtual Instrument

### 5. Conclusions

The main contribution of this paper lays in the proof-of-concept for wireless sensor networks integration with virtual instrumentation (LabVIEW) through dedicated software drivers. Therefore, we have developed an application which applies this to a Smart House monitoring and control scenario. Our approach is easily applicable to other WSN application domains such as generic environmental monitoring, energy monitoring and infrastucture and asset tracking. Experimental results have shown that, by using the XMesh networking protocol the system supplies data in a reliable manner to the base station and our virtual instrumentation project works as designed. The virtual instrumentation project we have developed has the potential to operate as a centralized framework to gather data from a variety of sources.

Envisioned future work includes the implementing a MDA300 data acquisition board driver in the LabVIEW XMesh WSN suite which is currently lacking. The MDA300 is a generic data acquisition expansion board for the IRIS platform. It offers analog input channels, digital input and output channels, relays and external sensor excitation. This opens up a whole range of new applications such as remote process control. Also, we are investigating the idea to develop a custom hardware design for the integration of specific sensors to the 51-pin connector on the IRIS radio/processing module. Our idea is to build an embedded networked platform using low-cost gas sensors for air monitoring such as: carbon dioxide ($CO_2$), carbon monoxide (CO), ozone ($O_3$) and particulate matter (PM).

R E F E R E N C E S

[1]. C. *Gomez, J. Paradells, J. Caballero,* Sensors Everywhere: Wireless Network Technologies and Solutions, F.V.E., 2010.

[2]. *M. A. Moisescu, I. S. Sacală, A. M. Stănescu*, Towards the Development of Internet of Things Oriented Intelligent Systems, U.P.B. Sci. Bull., ISSN 1454-234x, Series C, Vol. 72, Iss. 4, 2010.
[3]. *E. C. Rada, M. Ragazzi, M. Brini, L. Marmo, P. Zambelli, M. Chelodi, M. Ciolli*, Perspectives of Low-cost Sensors Adoption for Air Quality Monitoring, U.P.B. Sci. Bull., Series D, Vol. 74, Iss. 2, 2012.
[4]. *V. Sgârciu, G. Stamatescu, A. Năstase*, Optimizarea consumului energetic al unui Smart House folosind mediul de dezvoltare LabVIEW i tehnologii WSN (Energy Consumption Optimization of a Smart House using the LabVIEW Environment and WSN Technologies), Automatizări si Instrumentatie, Nr. 5-6, 2010.
[5]. *V. Sgârciu, G. Stamatescu*, Distance Process Monitoring using LabVIEW Environment, in LabVIEW – Modeling, Programming and Simulations (Ed. Riccardo de Asmundis), InTech, 2011, ISBN 978-953-307-521-210
[6]. *C. Volosencu, V. Malita*, Application of Virtual Instrumentation for Sensor Network Monitoring, 2nd International Conference on Manufacturing Engineering, Quality and Production Systems, 2010.
[7]. *National Instruments*, Crossbow XMesh WSN - Serial Driver for LabVIEW, 2007.
[8]. *O. Al-Khalid, et al.*, Wireless Sensor Networks for Swiftlet Farms Monitoring, World Academy of Science, Engineering and Technology, 60/2005.
[9]. *Crossbow Inc.*, MPR-MIB Users Manual, Rev. A, June 2007.
[10]. *Crossbow Inc.*, MTS/MDA Sensor Board Users Manual, Rev. A, June 2007.
[11]. *Crossbow Inc.*, XMesh User's Manual, Rev. C, March 2007.
[12]. *Sensirion*, SHT1X Humidity and Temperature Sensor IC Datasheet, 2012.